# A Robust Site Selection Model under uncertainty for Emergency and Disaster Management unit in Hong Kong

[1.] ***Heydari Mohammad.*** *Business College, Southwest University, Chongqing 400715, China; (Email: MohammadHeydari1992@yahoo.com)*
[2.] ***Fan Yanan.*** *Faculty of Earth Sciences and Environmental Management, University of Wrocław, Wrocław, Poland; (Email: 15735925764@163.com)*
[3.] ***Lai Kin Keung.*** *International Business School, Shaanxi Normal University, Xi'an 710062, China; (Email: mskklai@outlook.com)*

**Corresponding author**: International Business School, Shaanxi Normal University, Xi'an 710062, China. (E-mail: mskklai@outlook.com)

**Abstract**: This paper proposes two robust models for site selection problems for one of the major Hospitals in Hong Kong. Three parameters, namely, level of uncertainty, infeasibility tolerance as well as the level of reliability, are incorporated. Then, 2 kinds of uncertainty; that is, the symmetric and bounded uncertainties have been investigated. Therefore, the issue of scheduling under uncertainty has been considered wherein unknown problem factors could be illustrated via a given probability distribution function. In this regard, Lin, Janak, and Floudas (2004) introduced one of the newly developed strong optimisation protocols. Hence, computers as well as the chemical engineering [1069–1085] has been developed for considering uncertainty illustrated through a given probability distribution. Finally, our accurate optimisation protocol has been on the basis of a min-max framework and in a case of application to the (MILP) problems it produced a precise solution that has immunity to uncertain data.

## 1. Introduction

Researchers generally believe that the key to effectively ensure the supply of emergency supplies lies in rational location Allocation Problem (LAP) and scientific planning of the Vehicle Routing Problem (VRP). Moreover, there is an interdependent and interactional relationship between LAP and VRP. Thus, it is necessary to design and optimize them as a whole system, that is, to study Location-Routing Problem (LRP) in the emergency logistics system. LAP problem mainly refers to the decision-maker recognize the amount and location of facilities in a specific geographical area according to the geographical distribution of customers and the distribution of goods.

Hospital delivery throughout the world has been fraught with expensive, lower efficiency and poorer quality of the patients' care services and thus Hong Kong is no exception in this regard. However, qualities of the services provided for the patients have been far from exemplary. As an example, it is common that waiting time for specific routine surgeries at the public hospitals lasts 18 months. Another exacerbated issue would be the population aging; that is, the numbers of individuals aged ≤65 has been rapidly approached 15% of the general population in Hong Kong so that this figure has been enhancing by approximately 1% point annually.

Hospital under the present study is one of the major hospitals (MH) in Hong Kong under the management of the Hospital Authority (HA) and offers a full complement of services to Hong Kong people. MH provides medical care in both impatient and specialist outpatient services. However, the current constraints of QEH[1] are summarized below:

(I) Hospital with a high volume of activities;

(II) Reach the limits of capacity;

(III) Heavily utilized and aged infrastructure.

Moreover, in line with HA’s speech at the Hospital Authority Convention, 2013, the growing demand for hospital beds in Hong Kong is about 2600 by 2021, while about 6600 by 2031. Therefore, the long-term plan of government is to expand healthcare capacity. At this moment, MH is planning to open a new special hospital wards (SHW) in Hong Kong. The most important issue to determine the right site for this newly SHW.

The present paper is to present a robust site selection model for MH. The location of the hospital is critical in minimizing economic hardship for both MH and HA, and maximising the success of a Health Maintenance Organization (HMO) network since the location of the hospital definitely will have a direct effect on HMO utilization (Shuman et al., 1973). However, in the process of site selection, the cost, in terms of fixed and variable costs, and the government budget to operate the hospital, are uncertain, while unexpectedly increase the complexity of decision making.

[1] Queen Elizabeth Hospital

For addressing the issue of the parameter uncertainty, novel robust optimisation protocol approaches are developing. The robust optimisation protocol would handle the parameter uncertainty; supposing unknown factors belonged to a set of the bounded convex uncertainties. Therefore, it is possible to minimize the negative impacts on the objective values and guarantee the solutions (Düzgün & Thile, 2010). However, a basic concept behind the robust optimisation protocol would be the investigation of the worst scenario with any particular distribution assumption. Notably, Soyster (1973) has been the pioneer of the works on the robust optimisation protocol and required that each uncertain parameter reached its arrant case value that over-conserved the functional execution. Moreover, Ben-Tal, Nemirovski (1998, 1999, 2000) established a number of improvements with the ellipsoidal uncertainty set for adjusting conservatism levels and provided tractable mathematical reformulations. In addition, Bertsimas, Sim (2003; 2004) investigated uncertainties via describing a poly-hedron for all parameters. Consequently, they introduced the concept *"budget-of-uncertainty"* for controlling conservation. Furthermore, Düzgün and Thile (2010) demonstrated one range for all parameters would cause the overly conservative outcomes and illustrated uncertain parameters with numerous ranges. The other works followed a framework on the basis of scenario wherein uncertainty has been modelled by using some scenarios, which subsequently proposed the stochastic programming formulas (Mulvey et al., 1995). Nonetheless, it would be highly difficult to actually achieve precise probability data to distribute, which varied during times. Another study conducted via Lin et al. (2004) and Ben-Tal, Nemirovski (2000) developed a deterministic powerful alternative with regard to the amount of the data uncertainties, tolerance of in-feasibility, as well as levels of reliability as the probabilistic measurement has been applied. This protocol enjoys benefits like linearity, applicability. It also could be deterministically solved and easily controlled conservatism level. Therefore, the present robust optimisation protocol model has been designed with regard to the aforementioned worked in the field and specifically applied on the site selection model for MH.

## 2. Literature Review

Multiple works have been done on the production scheduling during the last years. A majority of the current works assumed that each datum has had known, fixed values. Nevertheless, uncertainty is really prevalent in a lot of scheduling problems, (Yang, T. H., et al., 2020) as a result of the absence of precise process models and variations in the procedure and environmental data. Therefore, new studies aimed to design techniques for addressing the scheduling problem based on un-certainty for creating reliable schedules that remained practicable in the exitance of the parameter uncertainty (refer to studies in Floudas (2005) as well as Floudas

& Lin (2004, 2005)). Hence, various methods could be utilized for the scheduling problem based on uncertainties like the probabilistic, stochastic as well as the fuzzy program (Sahinidis, 2004).

The study performed by Balasubramanian and Grossmann (2002) addressed the scheduling problem based on the demand's uncertainties. Researchers employed multi-stage stochastic MILP pattern wherein a number of decisions have been made by ignoring uncertainty. Moreover, other decisions have been made when uncertainty realized. Therefore, Balasubramanian and Grossmann introduced an approximation approach to solve some 2-phase models based on the shrinking horizon strategy. Additionally, Jia and Ierapetritou (2004) utilized concept of sensitivity analyses on the basis of inferences for the MILP problem for determining prominence of various limitations and factors in their scheduling system (Ozturk, C., & Ornek, M. A., 2016). This system provided a series of the candidate time-tables for ranging the unknown factors being considered. Furthermore, Bonfill, Bagajewicz, Espu˜na, and Puigjaner (2004) introduced a strategy for managing risks to schedule with uncertain demands. These researchers utilized a 2-phase stochastic optimisation pattern, maximising the predicted profits and managing the risks explicitly via studying a novel aim as the measure for controls, which caused a multi objective optimisation model. Then, the researchers (2005) dealt with the extension of the above model for considering unknown processing time. Therefore, a 2-phase stochastic procedure has been employed wherein a weighted sum of the predicted makes-pan and predicted waiting durations have been minimized; then, risks have been measured with various accuracy measures. Finally, (Ostrovsky, et al, 2004) dealt with extension of the 2-phase optimisation model for examining possible accurate estimating a number of unknown factors. Then, the split and bound strategy has been applied for solving the problem and has been on the basis of a partition of unknown area and approximation of the bounds in the objective function.

## 3. Robust optimisation protocol for General Integer and MLIP

In fact, the robust optimisation protocols have been designed for unknown information defined through multiple given distributions like normal distribution, smooth dispersion, difference of 2 normal distributions, binomial dispersion, general discrete distribution, as well as a Poisson dispersion. Actually, our powerful optimisation protocol introduced little auxiliary parameters and further limitations in the genuine MILP problem, which generated a deterministic strong counter-part problem, providing an optimal and possible solution with regard to relative volume of unknown information, level of reliability, as well as feasibility tolerance. Consequently, a powerful optimisation protocol has been utilized for the short-term scheduling issue based on uncertainties.

Now, general ILP with unknown factors would be studied. However, there is concern of the development of a strong optimisation protocol for generating

*"reliable"* solutions to ILP immune to the data uncertainty. Hence, a generic ILP with $m \times n$ parameters and $m$ constraint would be considered:

(1)

$$\begin{cases} \text{Max}\, cx \\ s.t.\, Ax \le b, x \ge 0, \end{cases}$$

So that $A$ refers to an $m \times n$ integer matrix of rank $m$ and $b \in \Re^m$. Moreover, uncertainty is the result of the left side variables of in-equality limitations, that is $b_i, i = 1, 2, ..., m$. However, there is concerns of the feasibility of the limitations below in a robust optimisation protocol framework:

(2)

$$\sum_{j \in J} a_{i,j} x_{i,j} \le b_i.$$

Based on Ben-Tal, Nemirovski (2000), when the nominal data have been partly worried, one or more constraints could be substantially violated. Therefore, this section aimed at the production of the accurate solutions to a generic ILP problem that had immunity to uncertainty. It is notable that our robust optimisation protocol has been initially provided via Ben-Tal, Nemirovski (2000) for LP problem, which has unknown coefficients and consequently Lin et al. (2004) extended it for addressing a MILP problem. Notably, our introduction method of robustness into the original model had a close similarity to the method employed in Lin et al. (2004). Hence, 2 kinds of uncertainty sets have been presented, including (I) bounded and (II) symmetric uncertainties.

### *3.1 Bounded uncertainty*

Let's assume that uncertainty data have been in a range as the following interval:

(3)

$$\left|\tilde{a}_{i,j} - \bar{a}_{i,j}\right| \le \varepsilon \left|\bar{a}_{i,j}\right|, \left|\tilde{b}_i - \bar{b}_i\right| \le \varepsilon \left|\bar{b}_i\right|,$$

where $\tilde{a}_{i,j}, \tilde{b}_i$ represent true values and $\bar{a}_{i,j}, \bar{b}_i$ refer to the nominal values. Moreover, $\varepsilon$ stands for the level of uncertainty.

The present study provided its own description of the precise solutions to ILP problem with the finite unknown left hand side variables:

**Definition 1.1.** When the uncertainty in ILP is illustrated in a bounded state, it is called solution $x$ correct if it meets this condition:

(i) $x$ would be possible for nominal problems,

(ii) if we have true values (say $\tilde{b}_i$ ) of uncertain parameters from intervals (3); therefore, $x$ should meet $i-th$ in-equality constraints with the error of Max $\delta \max\{1, |\bar{b}_i|\}$, wherein $\delta$ would be explained as a certain level of in-feasibility.

Particularly, condition (ii) could be written in this way:

$$\forall i(|\tilde{a}_{i,j} - \bar{a}_{i,j}| \leq \varepsilon |\bar{a}_{i,j}|, |\tilde{b}_i - \bar{b}_i| \leq \varepsilon |\bar{b}_i|,): \\ \sum_{j \in J} \bar{a}_{i,j} x_{i,j} + \varepsilon \sum_{j \in M_J} |\bar{a}_{i,j}| u_{i,j} \leq \tilde{b}_i + \delta \cdot \max\{1, |\bar{b}_i|\}. \tag{4}$$

Therefore, in order to derive the robust solution, the worst values of the unknown variables have been used:

$$\tilde{b}_i \geq \bar{b}_i - \varepsilon |\bar{b}_i|, \tag{5}$$

and substituted (5) into (4).

Hence, $x$ would be clearly correct if and just if $x$ would be one of possible solutions for optimisation problem below:

$$\begin{gathered} \max \quad cx \\ \text{s.t. } \sum_{j \in J} a_{i,j} x_{i,j} \leq \bar{b}_i \\ \sum_{j \in J} \bar{a}_{i,j} x_{i,j} + \varepsilon \sum_{j \in M_J} |\bar{a}_{i,j}| u_{i,j} \leq \bar{b}_i - \varepsilon |\bar{b}_i| + \delta \cdot \max\{1, |\bar{b}_i|\} \\ -u_{i,j} \leq x_{i,j} \leq u_{i,j} \\ x_{i,j} \geq 0, \forall i, j. \end{gathered} \tag{6}$$

In addition, the calculated formulation (6) would be known as " $(\varepsilon, \delta)$ -Interval Robust Counterpart ( $IRC[\varepsilon, \delta]$ )" of the original ILP issue.

### *3.2 Symmetric uncertainty*

Therefore, this sub-section supposed uncertain data $\tilde{a}_{i,j}, \tilde{b}_i$ as a random and symmetric distribution surrounding the nominal values $\bar{a}_{i,j}, \bar{b}_i$ as follow:

(7)

$$\tilde{a}_{i,j} = (1+\varepsilon\xi_{i,j})\bar{a}_{i,j}, \tilde{b}_i = (1+\varepsilon\varsigma_i)\bar{b}_i,$$

where the perturbations $\xi_{i,j}, \varsigma_i$ represent the independent variables with symmetric distribution in the interval $[-1,1]$.

In order to provide the same description of the correct solution to the ILP problem with the finite uncertainty, transferring a deterministic version (ii) to the common probabilistic one would be very crucial. Hence, we defined a correct solution to the ILP problem with the symmetric unknown variables:

**Definition 1.2.** When there is a symmetric uncertainty, the solution $x$ would be called correct if it meets this condition:

(i) $x$ would be possible for the nominal problems,

(ii) for each $i$, event probability of a limited violation; that is,

$$\sum_{j\in J} \tilde{a}_{i,j}\, x_{i,j} > \tilde{b}_i + \delta \cdot \max\{1, |\bar{b}_i|\},$$

Would be at most $\kappa$ so that $\delta > 0$ refers to a certain impossible tolerance and $\kappa > 0$ would be a certain level of reliability.

Hence, a correct situation to integer linear programming problem with the symmetric unknown parameters would be calculated via making solutions to ($\varepsilon, \delta, \kappa$)-robust counterpart (RC[$\varepsilon, \delta, \kappa$]):

(8)

$$\max \quad cx$$

$$\text{s.t.} \sum_{j\in J} \bar{a}_{i,j} x_{i,j} \le \bar{b}_i$$

$$\sum_{j\in J} \bar{a}_{i,j} x_{i,j} + \varepsilon \left[ \sum_{j\in M_J} |\bar{a}_{i,j}| u_{i,j} + \Omega \sqrt{\sum_{j\in M_J} \bar{a}^2_{i,j} v^2_{i,j} + \bar{b}^2_i} \right] \le \bar{b}_i + \delta \max\{1, |\bar{b}_i|\}$$

$$-u_{i,j} \le x_{i,j} - v_{i,j} \le u_{i,j}, \forall i, j,$$

$$x_{i,j} \ge 0, \forall i, j,$$

where $\Omega$ stands for a (+) parameter via $\kappa = \exp\{-\Omega^2 / 2\}$. Therefore, the case considered here has been a specific kind of the study conducted by Lin et al. (2004); hence, the procedure for deriving RC [ $\varepsilon, \delta, \kappa$ ] could be immediately called **Lemma 1** and **Theorem 2** reported by Lin et al.'s (2004) study that has been ignored in the present research for simplicity.

Notably, according to the above explanation of the relatively correct formulation, relative variables of uncertainties ( $\varepsilon$ ), reliability level ( $\kappa$ ), and in-feasibility tolerance ( $\delta$ ) have been supposed single and common for simplification. Nonetheless, these new robust optimisation protocols could be readily extended for considering more general cases where such parameters depended on the constraints.

## 4. Site Selection Model for QEH

The site selection model is concerned with choosing the right site for newly opened hospitals throughout a region such that the utilization is maximised. The involving model is related to three steps:

(i) Develop a utilization matrix;

(ii) Specify constraints;

(iii) Apply a robust optimisation protocol approach to select the best site.

### *4.1 Utilization Matrix*

For the purpose of locating a hospital optimally throughout a region, it is necessary to estimate the potential utilization or “attractiveness” of each choice. The mentioned utilization or “attractiveness” may be measured in terms of the number of subscribers who would select this site location under a dual-choice option.

The probability of the site selection as a function of travel time between the patient residence and hospital location and of socioeconomic attributes. The utilization function gives the expected utilization $u_{i,j}$ from population unit $i$ to facility location $j$ for each $i, j$ combination:

$$u_{i,j} = n_i p_{i,j}, \tag{9}$$

where $p_{i,j}$ is the probability of utilization for a person from population unit $i$ location $j$ as determined from the utilization function, and $n_i$ is the total population of unit $i$. The utilization matrix also gives the Max expected utilization for each choice:

(10)

$$u_j = \sum_{i=1}^{m} u_{i,j},$$

### *4.2 Constraints*

Constraints are imposed on the model to limit expenses, ensure that hospitals exceed a Min utilization, and assign each population unit to one and only one hospital, and restrict the number of hospitals.

The cost constraint could be considered in terms of fixed and variable costs. Fixed costs may represent set-up expenditures, including planning, capital outlays for equipment and facilities and recruiting the necessary core staff. Variable costs are those expenditures that are a function of the number of subscribers.

(11)

$$\sum_{j=1}^{n} f_j y_j + \sum_{j=1}^{n} v_j (\sum_{i=1}^{m} u_{i,j}) x_{i,j} \leq C,$$

where $f_j$ and $v_j$ are the fixed and variable costs, respectively, and $C$ represents the total amount of budget. The variables $y_j$ are 1 when a hospital is to be located at site $j$ and 0 otherwise.

It is believed that there exists a break-even point or Min utilization for a hospital.

(12)

$$\sum_{i=1}^{m} u_{i,j} x_{i,j} \geq k y_i,$$

So that $k$ is Min expected enrollment requirement before a hospital would be opened at the site $j$.

Each population unit would choose one and only one hospital, which could be mathematically stated as;

$$\sum_{j=1}^{n} x_{i,j} \geq 1. \tag{13}$$

Furthermore, the Max number of hospitals should be prescribed by HA managers,

$$\sum_{j=1}^{n} y_j \leq S. \tag{14}$$

Finally, the mathematical model to select the site to determine a location of new hospitals which maximises total utilization is presented below:

$$\text{Maximise } E = \sum_{i=1}^{m}\sum_{j=1}^{n} u_{i,j} x_{i,j} \tag{15}$$

$$\text{s.t. } \sum_{j=1}^{n} f_j y_j + \sum_{j=1}^{n} v_j (\sum_{i=1}^{m} u_{i,j}) x_{i,j} \leq C,$$

$$\sum_{i=1}^{m} u_{i,j} x_{i,j} \geq \mathrm{ky}_i,$$

$$\sum_{j=1}^{n} x_{i,j} \geq 1,$$

$$\sum_{j=1}^{n} y_j \leq S.$$

## 5. Robust Optimisation

It is possible to experience a direct application of the robust optimisation protocol models in Section 2 to the site selection problem with uncertain costs and budgets. Remembering the basic model (15), $IRC[\varepsilon,\delta]$ and $RC[\varepsilon,\delta,\kappa]$ application to this basic formulation has been considered. Considering the context of the hotel revenue management, parameters $\bar{b}_i$ in (6) and (8) would be surely positive. Therefore, it would be completely reasonable for assuming the existence of less than one booking request arriving every day. HENCE, $\max\{1,|\bar{b}_i|\} = \bar{b}_i$.

Consequently, the suggested robust site selection mannequins can be converted into the following formulations:

(16)

$$\text{Maximise } E=\sum_{i=1}^{m}\sum_{j=1}^{n}u_{i,j}x_{i,j}$$

$$\text{s.t. } \sum_{j=1}^{n}\overline{f}_j y_j+\sum_{j=1}^{n}\bar{v}_j(\sum_{i=1}^{m}u_{i,j})x_{i,j}\le\overline{C},$$

$$\sum_{j=1}^{n}\overline{f}_j y_j+\varepsilon\sum_{j\in M_J}\left|\overline{f}_j\right|s_j+\sum_{j\notin K_J}\bar{v}_j(\sum_{i=1}^{m}u_{i,j})x_{i,j}+$$

$$\sum_{j\in K_J}((\bar{v}_j+\varepsilon\left|\bar{v}_j\right|)(\sum_{i=1}^{m}u_{i,j})x_{i,j})\le(1-\varepsilon+\delta)\overline{C},$$

$$-s_j\le y_j\le s_j$$

$$\sum_{i=1}^{m}u_{i,j}x_{i,j}\ge \mathrm{k}\mathrm{y}_i,$$

(17)

$$\sum_{j=1}^{n}x_{i,j}\ge 1,$$

$$\sum_{j=1}^{n}y_j\le S.$$

and

$$\text{Maximise } E=\sum_{i=1}^{m}\sum_{j=1}^{n}u_{i,j}x_{i,j}$$

$$\text{s.t. } \sum_{j=1}^{n}\overline{f}_j y_j+\sum_{j=1}^{n}\bar{v}_j(\sum_{i=1}^{m}u_{i,j})x_{i,j}\le\overline{C},$$

$$\sum_{j=1}^{n}\overline{f}_j y_j + \sum_{j=1}^{n}\bar{v}_j(\sum_{i=1}^{m}u_{i,j})x_{i,j} +$$

$$\varepsilon\left[\sum_{j=1}^{n}\left|\overline{f}_j\right|l_j + \Omega\sqrt{\sum_{j\in M_J}\overline{f}_j^2 z_j^2 + \sum_{j\in M_J}\overline{v}_j^2 z_j^2 + \overline{C}^2}\right] \le (1+\delta)\overline{C},$$

$$\sum_{i=1}^{m}u_{i,j}x_{i,j} \ge \mathrm{ky}_i,$$

$$\sum_{j=1}^{n}x_{i,j} \ge 1,$$

$$\sum_{j=1}^{n}y_j \le S.$$

### *5.1 Robust optimisation protocol to schedule under uncertainty*

According to the research design, a robust optimisation protocol formulation[2] has been utilized to 4 instance problems. Each instance has been run through GAMS software (Brooke, et al., 2003) on a 3.20 GHz Linux work-station. Then, CPLEX 8.1 has been used to solve the MILP problems whereas DICOPT has been used to solve the MINLP (Viswanathan and Grossmann, 1990).

#### *5.1.1 Instance 1. Uncertainties via a Poisson distribution during the processing time*

Kondili, Pantelides, and Sargent (1993) initially designed an instance process utilized as a motivating instance in Section 1 of the article (Lin et al., 2004) on the finite uncertainties. Therefore, 2 products have been generated through 3 feeds in accordance with the State-Task Network (Figure 1). Then, STN utilized 3 various kinds of tasks that could be done in 4 diverse units. Table 1 reports the corresponding data for this instance that included suitability, capacity, processing time, as well as the storage limitation. It aimed at maximising the profit from selling the products fabricated in the time table equal to twelve hours.

Suppose that uncertainties during the processing time would have a Poisson distribution with the value equal to 5, the level of uncertainty ($\in$) equal to 5%, in0feasibility tolerance ($\delta$) equal to 20%, and the reliability level ($\kappa$) equal to 24% (relative to $\lambda$-value equal to 6). When *RC* [$\in$, $\delta$, $\kappa$] problem has been solved, a

[2] For more information about protocol formulation, see our other paper at (Heydari, M. et al., 2021).

*"robust"* schedule has been achieved (Figure 3) that investigated uncertainty in the processing time. Figure 2 depicts nominal schedule.

In comparison with nominal solution achieved at the nominal values of the processing duration, a correct situation exhibited much distinct scheduling approaches. As an instance, even the sequence of tasks in 2 reactors in Figure 3 experienced a significant deviation from Figure 1.

Moreover, in comparison with the nominal solution attained at the nominal values of the processing time, a correct situation exhibited so various scheduling approaches. As an instance, even the task sequences in 2 reactors in Figure 3 considerably deviated from

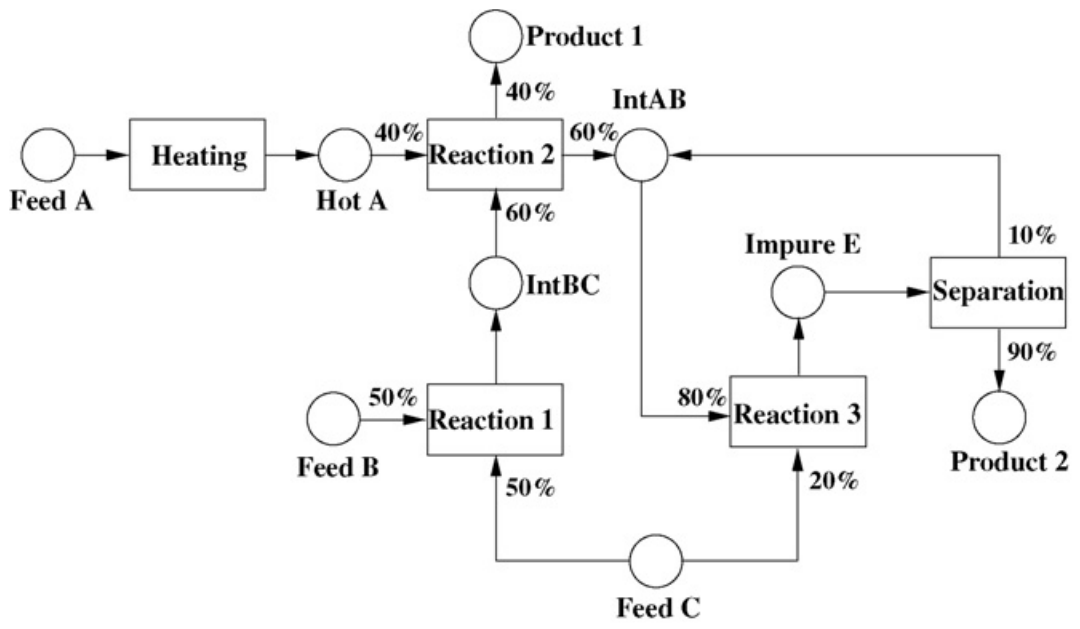

**Figure 1.** The state task network for instance 1.

**Table 1.** The data for instance 1.

| Units | Suitability | Processing time | Capacity |
|---|---|---|---|
| Heater | Heating | 1.0 | 100 |
| Reactor 1 | Reaction 1, 2, 3 | 2.0, 2.0, 1.0 | 50 |
| Reactor 2 | Reaction 1, 2, 3 | 2.0, 2.0, 1.0 | 80 |
| Separator | Separation | 2.0 | 200 |
| States | Initial Amount | Price | Storage |
| Feed A | Unlimited | 0.0 | Unlimited |
| Feed B | Unlimited | 0.0 | Unlimited |
| Feed C | Unlimited | 0.0 | Unlimited |
| Hot A | 0.0 | 0.0 | 100 |
| IntAB | 0.0 | 0.0 | 200 |
| IntBC | 0.0 | 0.0 | 150 |
| ImpureE | 0.0 | 0.0 | 200 |
| Product 1 | 0.0 | 10.0 | Unlimited |
| Product 2 | 0.0 | 10.0 | Unlimited |

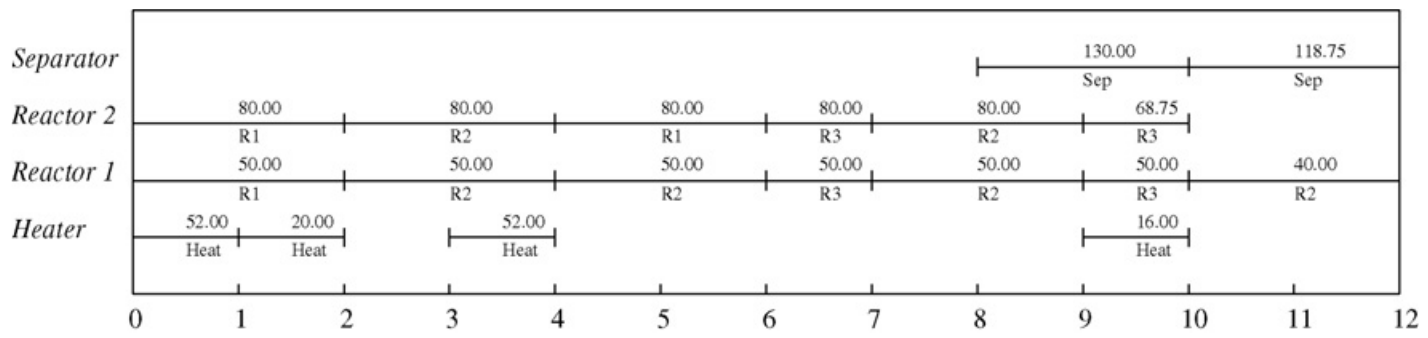

**Figure 2.** An optimal solution via the nominal time for processing (profit=3638.75).

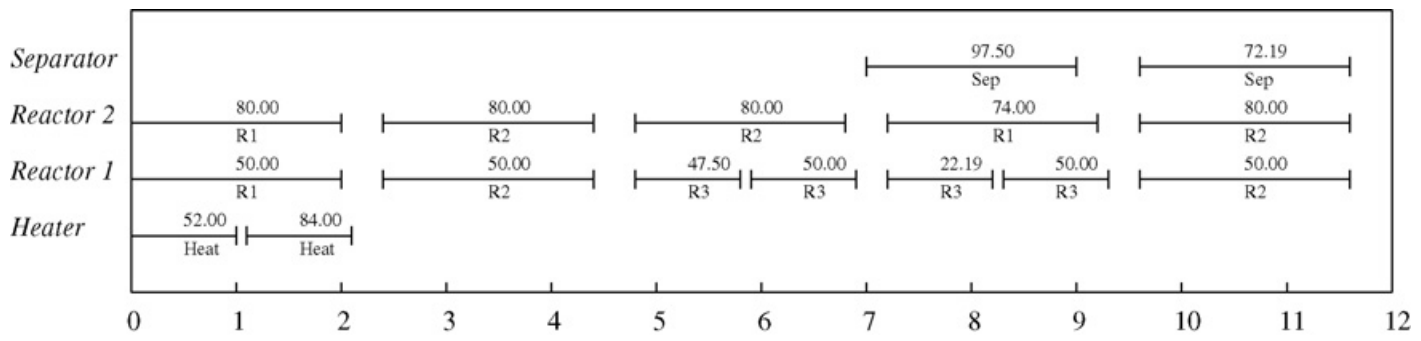

**Figure 3.** A robust solution via uncertain processing time (profit=2887.19).

Sequences in nominal solution in Figure 2. As seen, the correct solutions ensured feasibility of a correct schedule with a given level of uncertainty, level of reliability, and infeasibility tolerance. Nonetheless, reduction of the resultant profit from 3638.75 to 2887.19 has been reported, representing the effects of uncertainty on the overall production. Table 2 compares the pattern and solution statistic for robust and nominal solutions.

Figure 4 is a summary of the outputs of RC problem with numerous diverse mixes of infeasibility and uncertainty levels at enhancing values of the level of reliability. As shown, at a certain level of reliability, maximal profit, which could be gained, decreased by increasing the level of uncertainty, reflecting further conservative scheduling decisions due to uncertainty. Moreover, at the specific level of reliability, maximal profit increased by enhancing the tolerance level of infeasibility, meaning that it is possible to incorporate more aggressive scheduling arrangement if violations of the pertinent timing constraints could be further tolerated. Additionally, at the specific level of uncertainty and infeasibility tolerance, profit increased by enhancing the level of reliability, reflecting that probable violation of uncertain constraints allowed for more aggressive scheduling. Therefore, the obtained outputs would be compatible with intuition as well as other approaches. Nevertheless, considering the powerful optimisation protocol, impacts of uncertainties as well as trade-off between the opposed goals would be effectively and rigorously quantified.

**Table 2.** The mannequin and solution statistic for instance 1.

| | **Robust solution** | **Nominal solution** |
|---|---|---|
| Benefit | 2887.19 | 3638.75 |
| CPU time (s) | 11.33 | 0.46 |
| Binary variables | 96 | 96 |
| Continuous variables | 442 | 442 |
| Constraints | 777 | 553 |

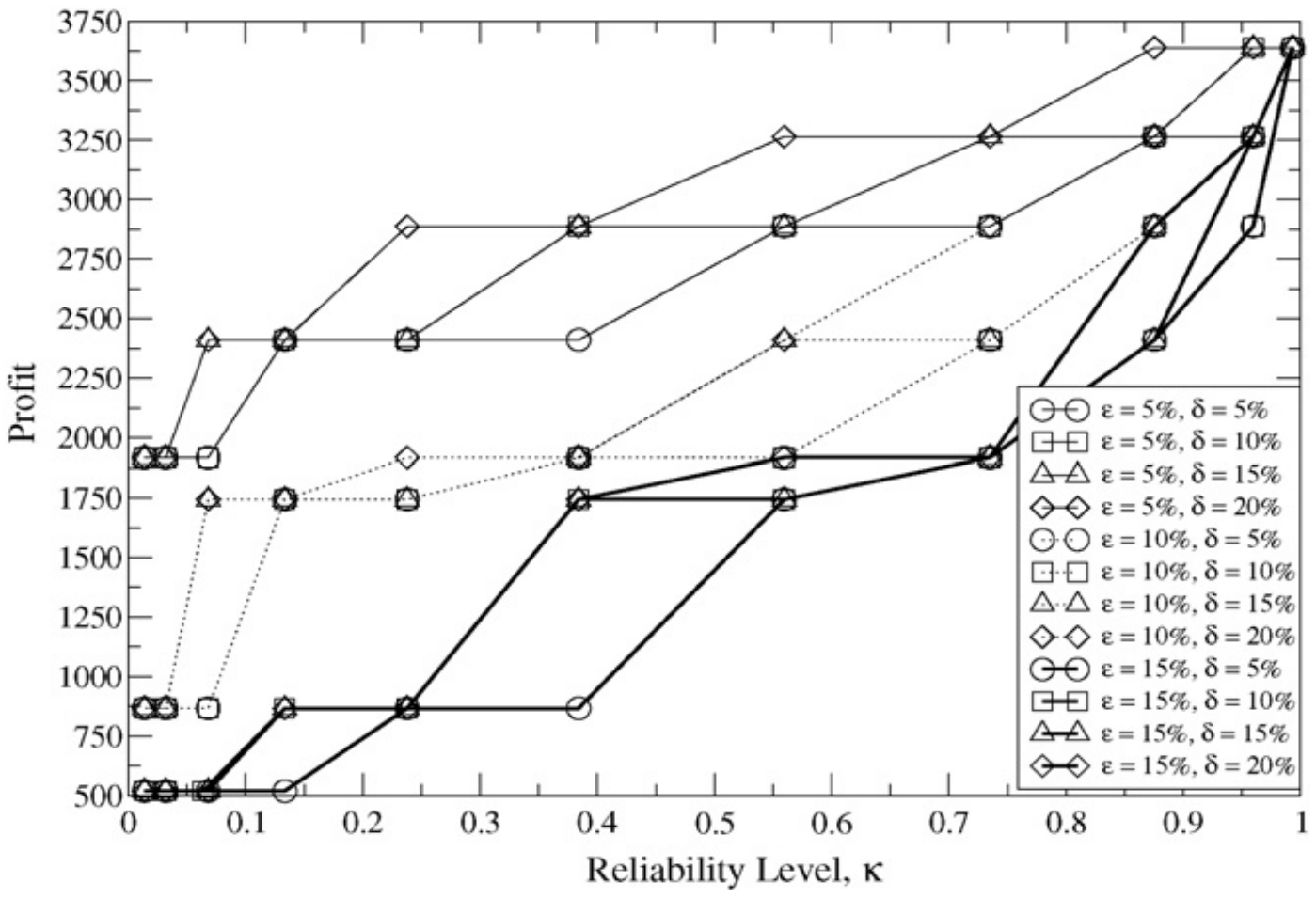

**Figure 4.** Level of profit versus reliability at different

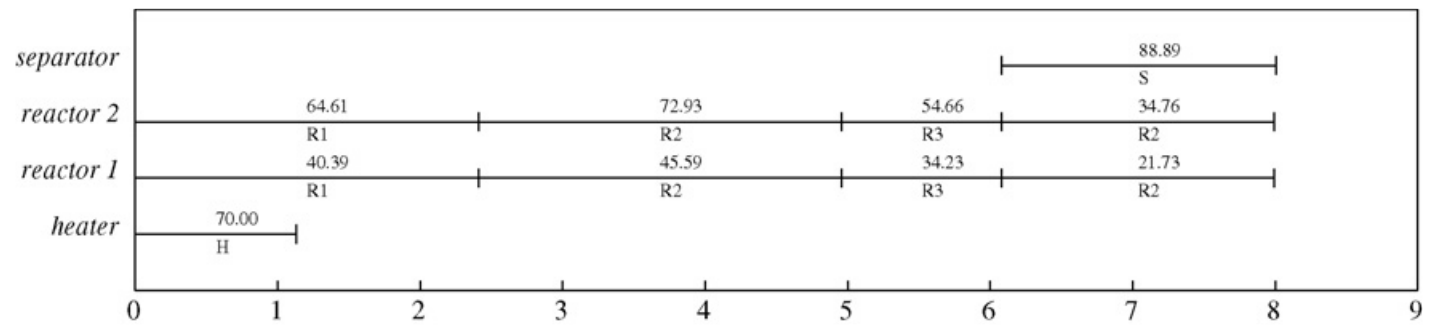

**Figure 5.** Optimum solution via the nominal goods demands (make-span=8.007).

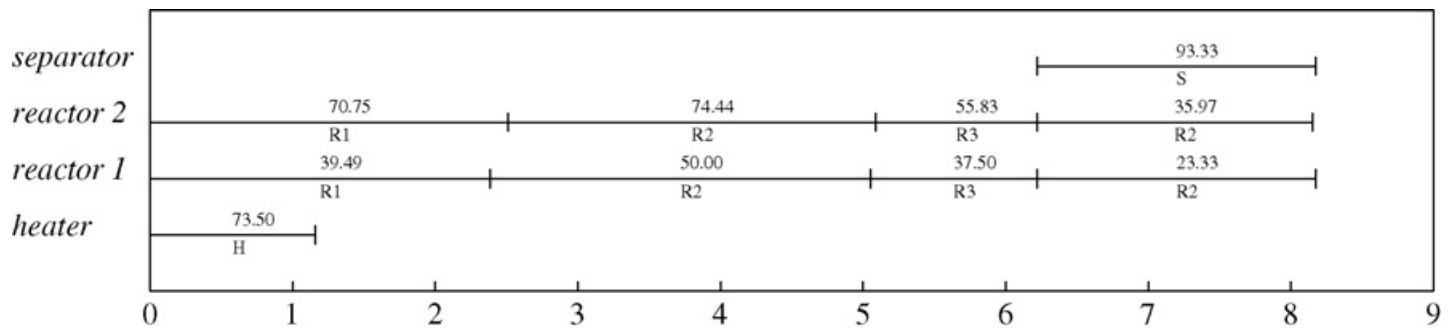

**Figure 6.** A robust solution via unknown goods demand (make-span=8.174).

#### *5.1.2 Instance 2. Uncertainty via a smooth distribution in the goods demand*

According to instance 2, uncertainties have been considered with a smooth dispersion in the goods demand for a similar procedure provided in instance 1. Nonetheless, the objective function has been to minimize the make-span for a certain demand equal to 70 for good 1 and 80 for good 2. Level of uncertainties $(\in)$ equaled 10% and in-feasibility tolerance (δ) equaled 5%. Moreover, level of reliability (κ) has been 0%. Figure 5 demonstrates nominal schedule with a make-span equal to 8.007. According to Figure 6, robust schedule has been achieved via solving the robust counterpart problem so that the corresponding make-span equaled 8.174. In case of the execution of the obtained schedule, make-span would be ensured to be at most 8.174 with a 100% probability in the exitance of 10% uncertainty in the product demand. Table 3 compares the pattern and solution statistic for robust and nominal solution.

Figure 7 represents a summary of the outputs of RC problem with multiple diverse combinations of infeasibility as well as uncertainty levels at the enhancing values of the level of reliability. As observed, at the certain level of reliability, Min make-span increased by enhancing level of uncertainty, indicating more conservative scheduling decisions, which further lasted due to uncertainties in the demand. Moreover, at the constant levels of reliability, Min make-span decreased by increasing the level of infeasibility tolerance, meaning that scheduling arrangements with higher aggression could be included if violation of the relevant demand limitations.

**Table 3.** The mannequin and solution statistic for instance 1.

| | **Robust solution** | **Nominal solution** |
|---|---|---|
| Make-span | 8.174 | 8.007 |
| CPU time (s) | 0.02 | 0.02 |
| Binary variables | 60 | 60 |
| Continuous variables | 280 | 280 |
| Constraints | 409 | 375 |

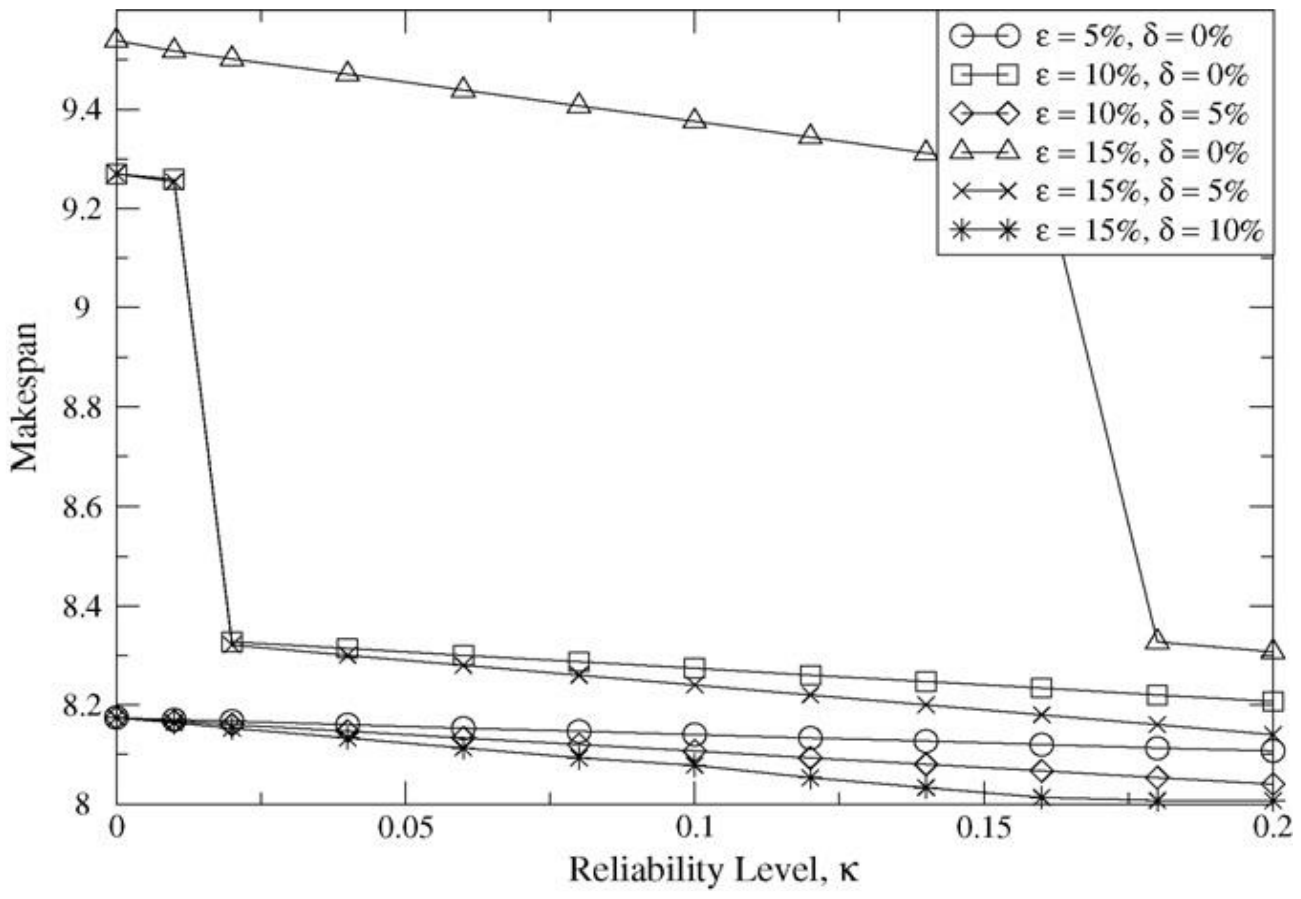

**Figure 7.** The make-span versus the level of reliability at diverse levels of uncertainties as well as in-feasibility level for instance 2.

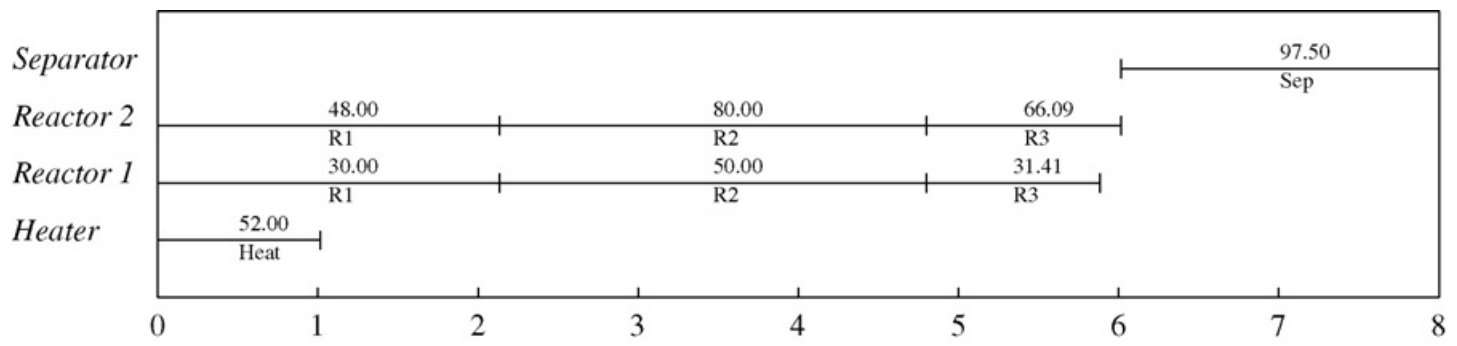

**Figure 8.** The optimised solution via the nominal market price (profits =1088.75).

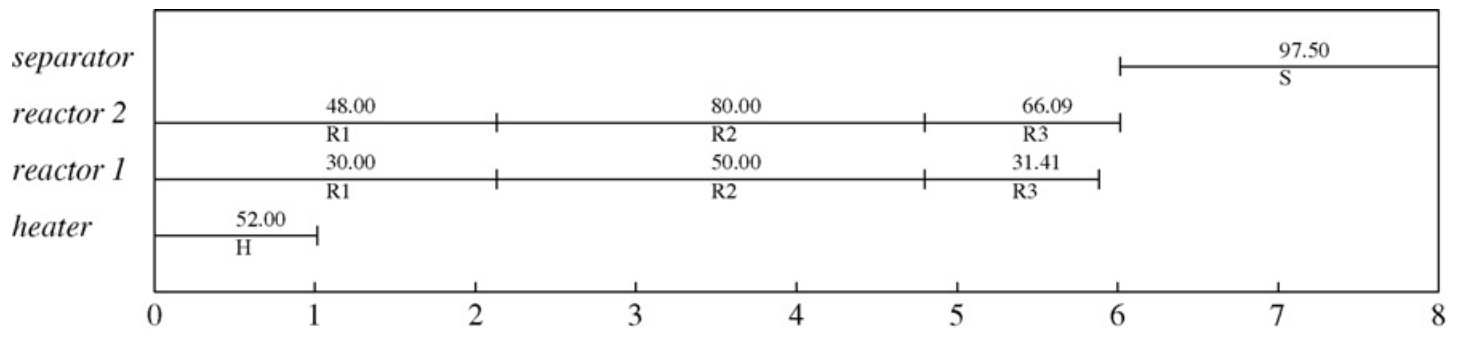

**Figure 9.** A correct solution via unknown market price (profits =966.97).

Could be further allowed. Put differently, at the certain level of uncertainty and infeasibility tolerance, make-span decreased by increasing the level of reliability, reflecting that probable violations of uncertain constraints allowed for more aggressive scheduling. Hence, it is possible to quantify the effects of uncertainty on schedule using a robust optimisation protocol.

### 5.1.3 Instance 3. Uncertainty via normal distribution in the market price

For this instance, level of uncertainty has been investigated via a normalized distribution in market price for a similar procedure in instance 1 and 2. Nonetheless, the objective function has been for maximising profits in eight hours. Level of uncertainty ($\in$), infeasibility tolerance (δ), and level of reliability (κ) equaled 5, 5, and 5%. Figure 8 shows a nominal schedule with the profit equal to 1088.75. Moreover, the robust schedule has been achieved via solving the robust counterpart issue (Figure 9) and the corresponding advantage equaled to 966.97. Upon the implementation of the schedule, the profit has been ensured to be not less than 966.97 with 95% probability in the exitance of 5% uncertainty in the raw materials and product prices. Table 4 compares the pattern and solution statistic for robust and nominal solution.

**Table 4.** The mannequin and solution statistic for instance 3.

| | **Robust solution** | **Nominal solution** |
|---|---|---|
| Benefit | 966.97 | 1088.75 |
| CPU time (s) | 0.05 | 0.02 |
| Binary variables | 60 | 60 |
| Continuous variables | 280 | 280 |
| Constraints | 334 | 334 |

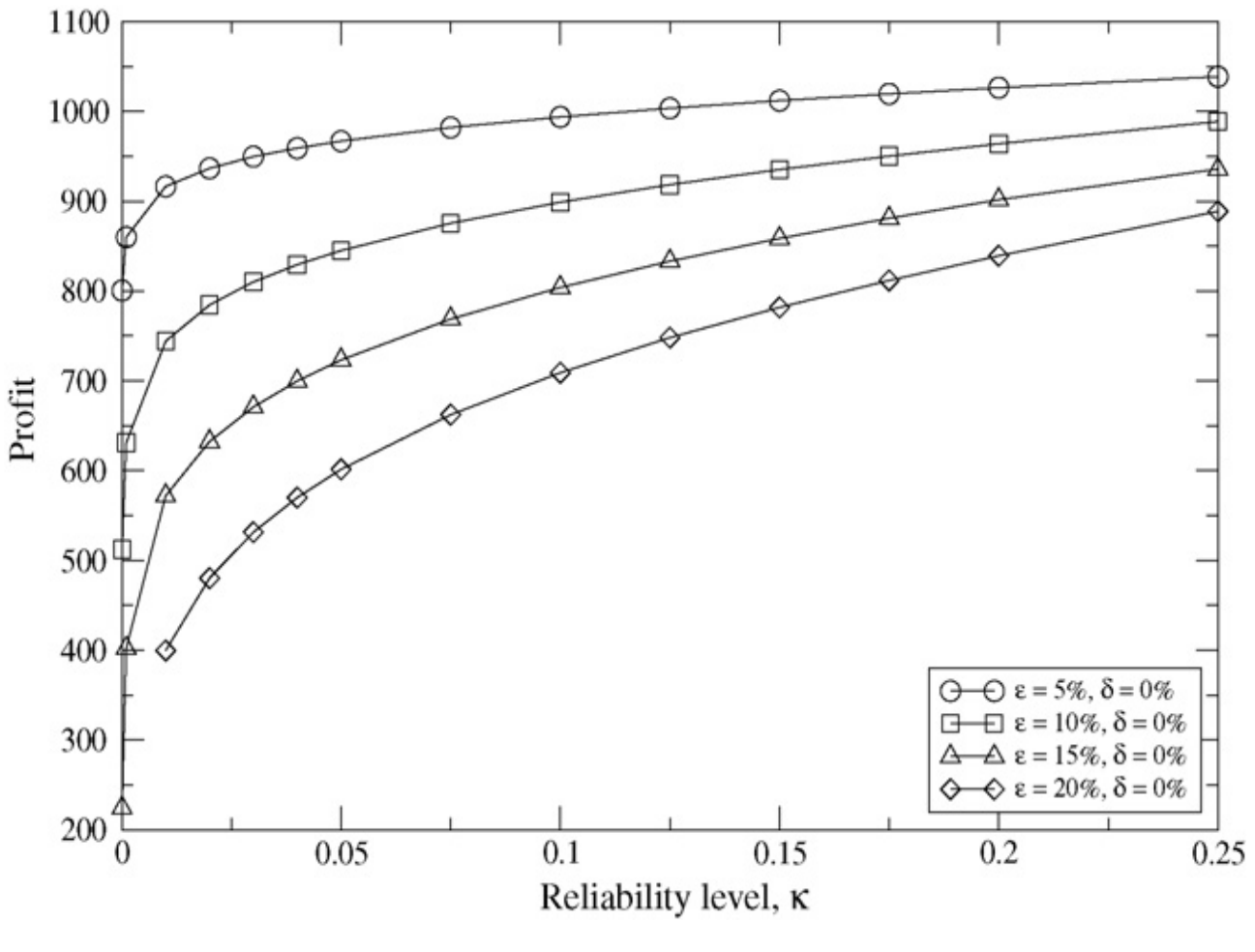

**Figure 10.** The profit versus level of reliability at diverse levels of uncertainties and in-feasibility for instance 3.

Figure 10 represents a summary of RC problem outputs at multiple diverse level of uncertainties and 0% in-feasibility tolerances at the enhancing amounts of level of reliability. As seen in the figure, at certain level of reliability, maximal profit, which could be gained, decreased by enhancing level of uncertainty, indicating more conservative scheduling decisions due to uncertainty. Moreover, at the certain level of uncertainty and infeasibility tolerance, profit increased by enhancing level of reliability, demonstrating that with the increase of probable violation of uncertain constraint or κ, λ decreased and the profit took on a greater value based on Equation (18):

$$Profit \le \sum_{s \in S_p} p_s . STF(s) - \in \lambda \sqrt{\sum_{s \in S_p} p_s^2 STF(s)^2 + \sum_{s \in S_r} p_s^2 STI(s)^2 + \delta} \tag{18}$$

So that $\lambda = F^{-1}(1-k)$ *and* $F_n^{-1}$ is reverse distribution functions of the random variables with the standard normalized distribution.

Notably, the above instance considered uncertainties during the procedure time of missions for the industrial case-report that has been initially provided in Lin et al.'s (2002) study. Therefore, actual plant data has been utilized for determining the kinds and levels of uncertainty in the processing time. In fact, industrial plant has been a multi-product chemical plant, which manufactured tens of various goods in accordance with a major 3-phase recipe and its changes with ten pieces of instrumentation. Therefore, the first sub-horizon in Lin et al.'s (2002) study contained 5 days and 8 various products. Hence, the objective function has been to maximise the general production described as the weighted sum of the substances aggregated at the conclusion of sub-horizon minus a penalty term for the lack of satisfaction of the demands with mid-term deadlines. Finally, a processing recipe demonstrated by Figure 11 has been utilized for each product.

This plant consisted of 3 kinds of units and each of them corresponded to 1 of 3 major processing operations. Therefore, 4 types of 1 unit (units 1 to 4) have been utilized for operation 1, 3 types of 2 units (units 5 to 7) have been employed for operation 2, and 3 types 3 units (units 8 to 10) have been utilized for operation 3. Then, type 1 units and type 3 units have been applied in the batch mode whereas type 2 units have been acted in a continual mode. Moreover, Table 5 presents nominal processing duration or the processing rates of all tasks in the relative proper units.

Therefore, for determining the forms of uncertainties in the processing duration or rate, we addressed analysis of the actual plant data. Hence, 2 various kinds of uncertainty have been selected on the basis of data; that is, uncertainties via a normal distribution and finite uncertainties. For bounded uncertainty, ranges for unknown variables have been provided, and mean and standard deviation (SD) for uncertain parameters have been determined for normal uncertainty. Moreover, a total number of 23 uncertain parameters has been recognized, including 8 in units 1 to 4, 5 in units 5 to 7, and 10 in units 8 to 10. Table 6 summarized characteristic nominal values, mean, range, and SDs for all uncertain parameters.

Now, strategy 2 for uncertainties in the processing duration or the rate in Section 3.2 would be utilized for the mentioned case-study. Besides major sequencing constraints, the processing times appeared in 2 further constraints associated with timing operation 1 missions:

(19)

$$T^s(i,j,n+1) \leq T^f(i,j,n) + H\left(2 - wv(i,j,n) - wv(i,j,n+1)\right)$$
$$\forall_i \in I_r, j \in J_i, n \in N, n \neq N,$$
$$T^s(i,j,n+1) \leq T^f(i,j,n) + tcl_{ii'} + H\left(2 - wv(i,j,n) - wv(i',j,n+1)\right)$$
$$\forall_i \in I_r, i, i' \in I_j, \mathrm{i} \neq \mathrm{i}', n \in N, n \neq N$$

So that $I_r$ represents a series of operation 1 tasks. $J_r$ refers to a collection of type 1 units appropriate for operation 1 tasks. When $T^f(i, j, n)$ variables substituted, further constraints have been proposed below for parameters via the bounded uncertainty for obtaining the robust counterpart issue:

(20)

$$T^s(i,j,n+1) - T^f(i,j,n) \leq a_{ij}^L.wv(i,j,n) +$$
$$\beta_{ij}^L.B(i,j,n) + H\left(2 - wv(i,j,n) - wv(i,j,n+1)\right) + \delta 2,$$

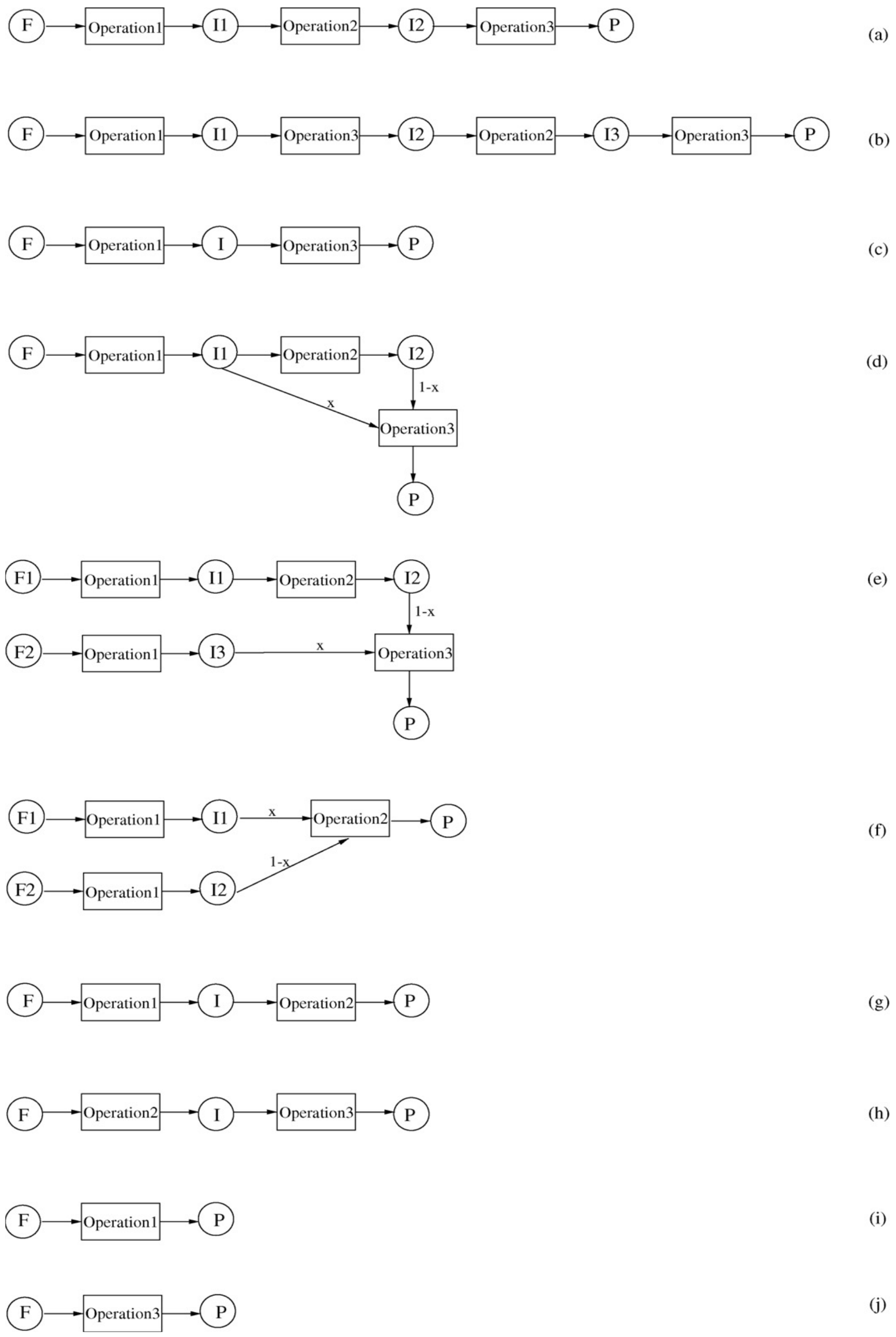

**Figure 11.** The state-mission network of the production recipe in industrial case-report.

**Table 5.** The nominal processing rates and times in an industrial case-report.

| | Units | | | | | | | | | |
|---|---|---|---|---|---|---|---|---|---|---|
| **Mission** | 1 | 2 | 3 | 4 | 5 | 6 | 7 | 8 | 9 | 10 |
| **1** | 0 | 0 | 0 | 9.5 | -- | -- | -- | -- | -- | -- |
| **2** | -- | -- | -- | -- | 0 | 0 | 0.95 | -- | -- | -- |
| **3** | -- | -- | -- | -- | -- | -- | -- | 12 | 12.8 | 12.5 |
| **4** | 0 | 10 | 10 | 10 | -- | -- | -- | -- | -- | -- |
| **5** | -- | -- | -- | -- | 0.575 | 0.575 | 0.725 | -- | -- | -- |
| **6** | -- | -- | -- | -- | -- | -- | -- | 12 | 12.8 | 12.5 |
| **7** | 6.09 | 6.09 | 6.09 | 11.1 | -- | -- | -- | -- | -- | -- |
| **8** | -- | -- | -- | -- | 0.6 | 0.6 | 0.8 | -- | -- | -- |
| **9** | -- | -- | -- | -- | -- | -- | -- | 12.5 | 13.8 | 12.9 |
| **10** | 6.09 | 6.09 | 6.09 | 11.1 | -- | -- | -- | -- | -- | -- |
| **11** | -- | -- | -- | -- | 0.6 | 0.6 | 0.8 | -- | -- | -- |
| **12** | -- | -- | -- | -- | -- | -- | -- | 12.5 | 13.8 | 12.9 |
| **13** | 6.09 | 6.09 | 6.09 | 11.1 | 0.6 | 0.6 | 0.8 | -- | -- | -- |
| **14** | -- | -- | -- | -- | -- | -- | -- | -- | -- | -- |
| **15** | -- | -- | -- | -- | -- | -- | -- | 12.5 | 13.8 | 12.9 |
| **16** | -- | -- | -- | -- | 0.6 | 0.6 | 0.8 | -- | -- | -- |
| **17** | -- | -- | -- | -- | -- | -- | -- | 12.5 | 13.8 | 12.9 |
| **18** | 0 | 8.5 | 8.5 | 0 | -- | -- | -- | -- | -- | -- |
| **19** | -- | -- | -- | -- | -- | -- | -- | 0 | 15 | 16 |
| **20** | 0 | 0 | 8.38 | 9.5 | -- | -- | -- | -- | -- | -- |

(21)

$$T^s\left(i,j,n+1\right)-T^s\left(i^{'},j,n\right)\leq a_{ij}^L.wv\left(i^{'},j,n\right)+$$
$$\beta_{ij}^L.B\left(i^{'},j,n\right)+tcl_{ii^{'}}+H\left(2-wv\left(i,j,n\right)-wv\left(i^{'},j,n+1\right)\right)+\delta 2,$$

Here $a_{ij}^L=\left(1-\in\right)a_{ij},\beta_{ij}^L=\left(1-\in\right)\beta_{ij}, and\ \delta 2$ represent a varied and correlate as followed with factor δ, which participated in further limitations relative to major sequencing limitations:

(22)

$$\delta+\delta 2=a^U-a^L=2.\in.\text{a},$$
$$or\ \delta+\delta 2=\beta^U-\beta^L=2.\in.\beta.$$

Accordingly, further limitations would be proposed for variables with normal uncertainties:

(23)

$$T^{s}(i,j,n+1)-T^{s}(i,j,n)\leq\left[1-\in\left(\lambda_{ij}^{a}.\sqrt{\sigma_{ij}^{a}}-\mu_{ij}^{a}\right)\right].a_{ij}.wv(i,j,n)$$
$$+\left[1-\in\left(\lambda_{ij}^{\beta}.\sqrt{\sigma_{ij}^{\beta}}-\mu_{ij}^{\beta}\right)\right].\beta_{ij}.\beta(i,j,n)$$
$$+H\left(2-wv(i,j,n)-wv(i,j,n+1)\right)+\delta 2,$$

(24)

$$T^{s}(i,j,n+1)-T^{s}(i^{'},j,n)\leq\left[1-\in\left(\lambda_{i^{'}j}^{a}.\sqrt{\sigma_{i^{'}j}^{a}}-\mu_{ij}^{a}\right)\right].a_{i^{'}j}.wv(i^{'},j,n)$$
$$+\left[1-\in\left(\lambda_{i^{'}j}^{\beta}.\sqrt{\sigma_{i^{'}j}^{\beta}}-\mu_{i^{'}j}^{\beta}\right)\right].\beta_{i^{'}j}.\beta(i^{'},j,n)+tcl_{i^{'}i}$$
$$+H\left(2-wv(i,j,n)-wv(i^{'},j,n+1)\right)+\delta 2$$

So that δ2 would be described as:

(25)

$$\delta+\delta 2=2.\in\left(\lambda.\sqrt{\delta}-\mu\right).$$

As seen, the objective function for the above issue would be to maximise the production with regard to the relative values of each state minus a penalty term for the lack of contentment of the demand at the intermediate due date:

(26)

$$\gamma\sum_{s}vald_{s}valp_{s}valm_{s}STF(s)-\sum_{s}\sum_{n}pri_{sn}SL(s,n)$$
$$\forall_{s}\in S,\ n\in N$$

**Table 6.** The bounded and normalized uncertainties in the processing duration and rate for the case-report.

| Mission | Unit | Nominal Value | Uncertainty | Rang | Mean | S.D. |
|---|---|---|---|---|---|---|
| 1 | 4 | 9.5 | Normal | -- | 9.912 | 0.523 |
| 7,10,13 | 1-3 | 6.09 | Normal | -- | 6.153 | 0.152 |
| 7,10,13 | 4 | 11.1 | Bounded | 10.1-11.3 | -- | -- |
| 20 | 3 | 8.38 | Bounded | 8.00-10.42 | -- | -- |
| 2 | 7 | 0.95 | Normal | -- | 0.9611 | 0.112 |
| 8,11,14,16 | 5-6 | 0.60 | Bounded | 0.344-0.853 | -- | -- |
| 3,6 | 9 | 12.8 | Bounded | 10.5-19.3 | -- | -- |
| 9,12,15,17 | 9 | 13.8 | Bounded | 12.0-16.3 | -- | -- |
| 9,12,15,17 | 10 | 12.9 | Normal | -- | 12.100 | 0.760 |

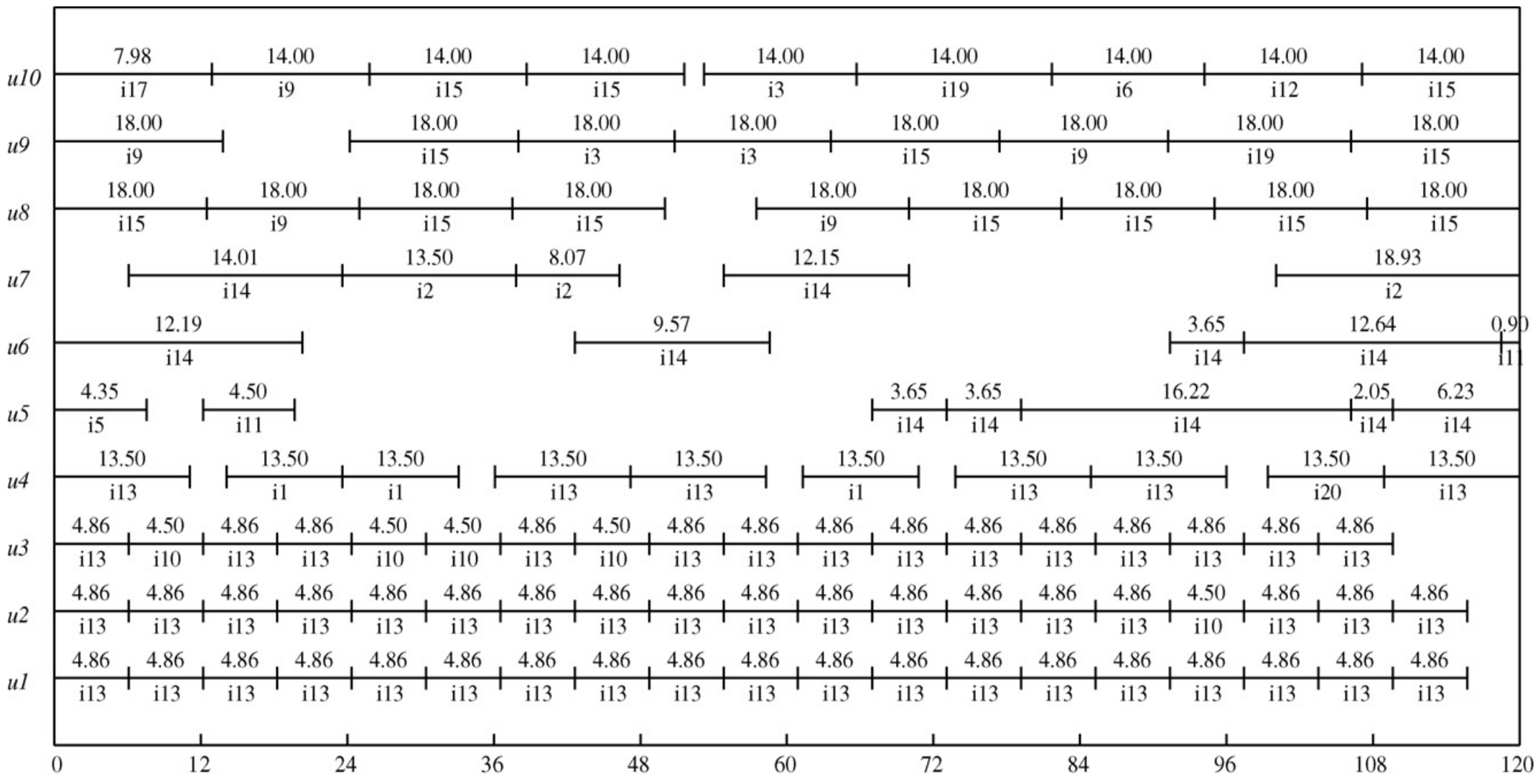

**Figure 12.** The nominal in the case-report schedule (Objective (Obj))=121.37, U1–U10; units I1 to I19: missions).

So that $vald_s$ represents the corresponding values of the relative product reflecting the respective significance for fulfilling the future demands. In addition, $valp_s$ refers to the relative value of the corresponding products representing the respective priority and $valm_s$ stands for the relative values of the state (s) in the materials sequences for the corresponding products. Moreover, STF(s) indicates the amounts of state (s) at the end of horizon and $pri_{sn}$ refers to the demand priority for state (s) at the event point (n). Furthermore, SL (s, n) represents a slack variable for the number of state (s), which has not met the demand at the event point (n) and γ stands for a fixed coefficient applied for balancing the relative value of 2 terms in the objective function.

It should be noted that this problem required additional sequencing constraints (19) to (21) for accurate scheduling of operation 1 tasks. However, the use of such constraints for describing the problem uncertainty led to a (MILP) problem, which has been caused by the situation that such uncertain constraints contained just 1 uncertain parameter, causing the linear deterministic kinds for normal unknown limitations and bounded limitations.

### *5.2 Computational outputs and discussion*

Figure 12 depicts a nominal solution for the above problem with continual time formula and objective function values equaled 121.37. Figure 13 presents a solution to the correct respective problem with each 23 uncertain parameter at 10% (relative) in-feasibility tolerances ($\delta$) for the finite unknown variables, 20% normal uncertain parameters 5% level of uncertainty ($\in$), and 5% level of reliability ($\kappa$) for normal uncertain parameters, reflecting just 5% violation of the constraints. As seen, objective function value has been 105.76 and the processing time of all uncertain tasks has been extended for ensuring that the schedule would be feasible at the given level of uncertainty, level of reliability, and in-feasibility tolerances. Nonetheless, value of the objective function declined. Moreover, a more precise investigation of the term related to the Obj function indicated declining the Obj function value for a correct solution as the relative value of violating intermediate due-dates enhanced whereas overall production diminished. Table 7 compares pattern and solution statistic for correct and nominal solution of an industrial case-study. The published CPU times indicated the time lasted for obtaining the most acceptable solution inside a time limit of 2 h.

**Table 7.** The mannequin and solution statistics for the case-report.

| | **Robust solution** | **Nominal solution** |
|---|---|---|
| Obj | 105.76 | 121.37 |
| Binary variables | 930 | 930 |
| Continuous variables | 6161 | 6005 |
| Constraints | 22931 | 18907 |
| CPU time (s) | 5910 | 3880 |
| Nodes | 35640 | 15230 |

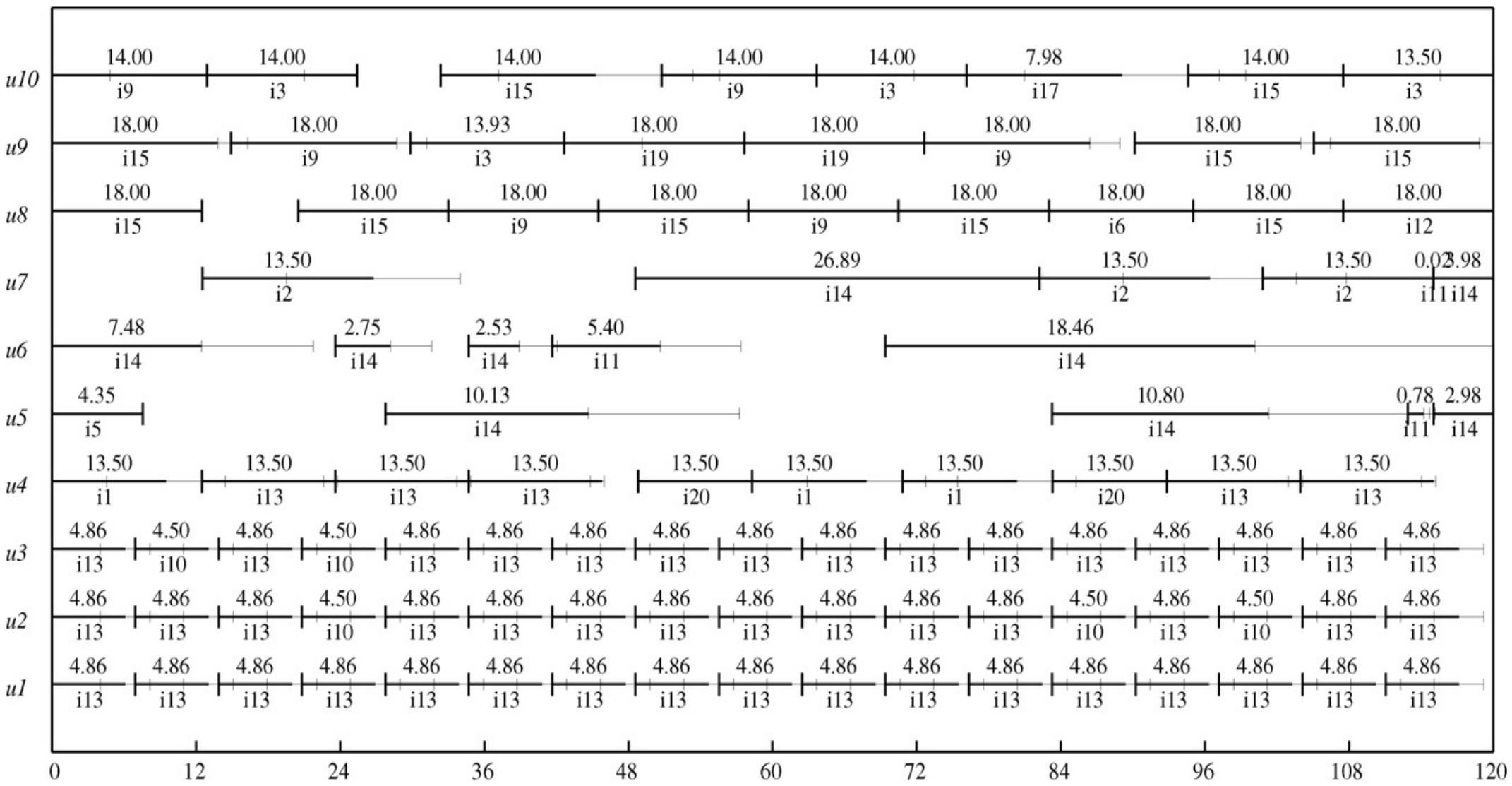

**Figure 13.** The correct time in a case-study (Obj=105.76, δ=10%, 20%, U1–U10, units I1 to I19, missions. Note: End-point of all thick horizontal lines indicated the task end time on the basis of nominal processing duration whereas thin vertical lines represented the end time range because of uncertainty in the processing time.)

It is notable that for a short-term timing problem examined in Part 5.1.1, the conservatism level observed in resultant production time-table varied with the distribution function of probability utilized for describing uncertain parameters. Comparing all types of uncertainties on each 3 uncertain constraint with diverse problem parameters for the mentioned instance problem indicated that (i) general discrete distribution has been constantly the least conservative that gave the most reasonable Obj function values, (ii) bi-nomial and Poisson distributions constantly gave the most conservative outputs, which yield the arrant Obj function values, (iii) among other kinds of distributions; that is, bounded, uniform, and normal, normal distribution has been commonly more conservative than the other 2 ones and smooth dispersion would be generally less conservative in comparison to the other two ones. Nonetheless, relative outputs for the 3 distributions strongly depended on the value of level of reliability, κ. Therefore, when κ increased, which indicted greater reliability, normal distribution became less conservative in comparison to the other two ones. Actually, for great κ, normal distribution would be less conservation in comparison to the bounded distribution.

Additionally, even though it is possible to use the robust optimisation protocol formulation for modelling the uncertainty in diverse MILP problems, a number of constraints of our approach. As an example, a series of probability distribution functions could be just utilized for limitations containing a single uncertain variable; that is, uniform, binomial, and Poisson. Such a condition would be caused by limitations in the probability theory and not the introduced formulation. In addition, it has been not possible for the robust optimisation protocol formulation[3] to examine the

[3] ***Notice:*** In each hospital (Usually a major trauma center), we have a red or black alert designed for immediate response to major incidents. During the major incident for dealing with mass casualties or major incidents, the hospital follows a one hundred ninety full pages protocol which called (major incident plan) stand by means prepare to receive casualties which we define as follows;

When we start implementing this protocol, we need more help for immediate response to the catastrophic incident, like even off-duty surgical residents from all wards. For example, ambulance services only take critically ill patients into the emergency department or casualty (A&E).

***Major incident situation happened, how CCUs, and specialist staff act?*** Between (5%-15%) of patients presenting to hospital following a bomb blast or terrorist-associated mass casualty incident will need intensive care (Adam, S. et al., 2017). The demands on critical care sources for other incidents, like large fires or natural disasters, are less well recorded (Avidan et al. 2007). However, following the aftermath of Hurricane Katrina in New Orleans, the Charity Hospital had to manage enhanced demand without the opportunity to evacuate for many days (Rubinson et al. 2008).

In 2007 the US Taskforce for Mass Critical Care offered that CCU plan to represent emergency mass critical care at three times the existing capacity for up to ten days (de Boisblanc et al. 2005). However, CCU beds and specialists/staff are a limited source usually fully utilised; in a major incident, releasing beds or expanding the source needs good planning and organisation. As critical care sources will often represent the major limiting element encountering large numbers of casualties, it is also essential that early connections with different hospitals be made to transfer more stable patients in the most secure way possible. Based on this, critical care managers/leaders must be involved in planning for mass casualties along the healthcare communities.

The CHEST Task Force for Mass Critical Care has suggested different levels of capacity expansion needs via different levels of casualty (Christian et al. 2014).

Critical care sources require to enhance for:

(a) Conventional response-able to expand immediately via at least 20% above baseline incentive care unit Max capacity;

(b) Crisis response-able to expand via at least 200% above baseline incentive care unit Max capacity via regional, local, national, and international agencies;

(c) Contingency response-able to expand rapidly via at least 100% above baseline incentive care unit (ICU) Max capacity via accessing local and regional resources;

The Min requirements for critical care offered via the EMCC taskforce (Rubinson et al. 2008) are:

- Vasopressor administration

dependent uncertain parameters that have been associated by general non-linear expressions; however, it could be used for linear dependence on the uncertain

---

- Mechanical ventilation
- Sedation and analgesia
- Optimal therapeutics and interventions, like renal replacement therapy and nutrition for those patients who are not able to take food via mouth, if warranted via the hospital or regional preference
- IV fluid resuscitation
- Antidote or antimicrobial administration for special disease processes, if applicable
- Algorithms to decrease adverse consequences of critical care and critical illness.

It is offered that a tiered response shall be set up, allowing increasingly more high-risk management practices to be viewed as the effect of the incident enhances. Ultimately, triage of patients for accessibility to scarce critical care sources would also be included.

Predict and manage such an incident contains:

- Training and education of specialist/staff
- A level of stockpiling or recognition of alternative resources of equipment (e.g., use of anaesthetic ventilators to supply ventilatory support, and NIV machines)
- Recognition of Specialist/staff via transferable skills like recovery nurses, respiratory nurses and previous critical care nurses;

When people still coming in but there is no way to put them out, we go to the red alert; In this condition basically, the number of coming in patents is excessed the number going out, and there is nowhere to put them, and we haven't got any spare capacity (bed), we get to the point where we are genuinely thinking can we treat these sick people? A red alert demonstrates there is a significant risk to the patients safety. In this condition, all non-emergency surgeries should be canceled, and we need to search for a free bed to start to discharge the beds until the flow moves forward. Whenever the hospital is on the alert, all wards are under pressure to discharge the patients. What we require to do is follow the patients from inter to leave.

When all wards are full, and there is no bed in any wards; at the moment, the emergency admission system is running at its absolute limit; it is starched. The point where if one person, one key person doesn't turn on for whatever the reason, then whole things just can fall apart. Just like a real domino, just one person comes out the whole thing just collapses. Because the flow stops, you get blockage and back right away up to the front door. Everything just becomes impossible. Whenever the hospital is on alert, all of the wards are under pressure to discharge the patients (Heydari, M. et al., 2021).

parameters. Ultimately, our formulation could control the uncertainties in the linear constraints; therefore, further studies should address such issues.

## 6. Conclusion

The present research investigates the site selection problem for MH in Hong Kong. A robust optimisation protocol algorithm is employed to deal via uncertain parameters in a mathematical model. Two scenarios of uncertainty have been investigated. Therefore, the current study proposed a novel method for addressing timing based on the problem of uncertainties on the basis of the correct optimisation protocol so that in a case of application to the MILP problems produced a *"robust"* solution, which had immunity against uncertainty in coefficient and the right-side variables of in-equality limitations. This strategy could be utilized for addressing issue of production scheduling with market demands, unknown processing duration, and the product as well as raw materials price (Ierapetritou, Floudas, 1998a, 1998b; Lin, Floudas, 2001). Moreover, the new computational outputs showed that our new method provided one of the efficient ways for addressing the timing difficulties based on un-certainty, which produced reliable time-tables and generated beneficial information of the tradeoffs between the inconsistent goals. Consequently, as a result of its efficacious transformation, this approach could solve the real-world problems with multiple uncertain parameters (Janak et al., 2004, Lin et al., 2002, 2003; 2006a, 2006b).